\documentclass[onecolumn]{quantumarticle}
\usepackage{amsmath}
\usepackage[numbers]{natbib}
\usepackage{amssymb}
\usepackage{amsthm}
\usepackage{amsfonts}
\usepackage[caption=false]{subfig}
\usepackage[colorlinks]{hyperref}
\usepackage[all]{hypcap}
\usepackage{tikz}
\usepackage[section]{placeins}
\usepackage{listings}
\usepackage{color,soul}
\hypersetup{colorlinks,citecolor=blue,urlcolor=blue,linkcolor=blue}
\usepackage[utf8]{inputenc}
\usepackage{capt-of}
\usepackage{multirow}

\DeclareFixedFont{\ttb}{T1}{txtt}{bx}{n}{12} 
\DeclareFixedFont{\ttm}{T1}{txtt}{m}{n}{12}  

\usepackage{color}
\definecolor{deepblue}{rgb}{0,0,0.5}
\definecolor{deepred}{rgb}{0.6,0,0}
\definecolor{deepgreen}{rgb}{0,0.5,0}

\newcommand\pythonstyle{\lstset{
language=Python,
basicstyle=\ttm,
otherkeywords={self,controlled_by,with,Quint,LookupTable},
keywordstyle=\ttb\color{deepblue},
emph={measure,__init__},
emphstyle=\ttb\color{deepblue},
stringstyle=\color{deepgreen},
showstringspaces=false
}}
\lstnewenvironment{python}[1][]
{
\pythonstyle
\lstset{#1}
}
{}

\newcommand\qsharpstyle{\lstset{
language=C,
basicstyle=\ttm,
otherkeywords={self,controlled_by,with,Quint,LookupTable},
keywordstyle=\ttb\color{deepblue},
emph={body,Adjoint,auto,adjoint,let,if,else,operation,LittleEndian,Unit,Length,for,in},
emphstyle=\ttb\color{deepblue},
stringstyle=\color{deepgreen},
showstringspaces=false
}}
\lstnewenvironment{qsharp}[1][]
{
\qsharpstyle
\lstset{#1}
}
{}

\newcommand{\eq}[1]{Eq.~\hyperref[eq:#1]{(\ref*{eq:#1})}}
\renewcommand{\sec}[1]{\hyperref[sec:#1]{Section~\ref*{sec:#1}}}
\DeclareRobustCommand{\app}[1]{\hyperref[app:#1]{Appendix~\ref*{app:#1}}}
\newcommand{\fig}[1]{\hyperref[fig:#1]{Figure~\ref*{fig:#1}}}

\begin{document}

\title{Asymptotically Efficient Quantum Karatsuba Multiplication}

\date{\today}
\author{Craig Gidney}
\email{craiggidney@google.com}
\affiliation{Google Inc., Santa Barbara, California 93117, USA}

\begin{abstract}
We improve the space complexity of Karatsuba multiplication on a quantum computer from $O(n^{1.427})$ to $O(n)$ while maintaining $O(n^{\lg 3})$ gate complexity.
We achieve this by ensuring recursive calls can add their outputs directly into subsections of the output register.
This avoids the need to store, and uncompute, intermediate results.
This optimization, which is analogous to classical tail-call optimization, should be applicable to a wide range of recursive quantum algorithms.
\end{abstract}

\maketitle

\section{Introduction}
\label{sec:introduction}

A quantum computer can evaluate any circuit that a classical computer can evaluate.
Given this fact, it is somewhat surprising that porting classical algorithms to work on a quantum computer is not a trivial task.
The issue is that classical algorithms and circuits perform irreversible operations (e.g. discarding initialized memory, looping until a condition is satisfied).
This prevents their use as subroutines in larger quantum algorithms, such as Grover search, because irreversible operations cause decoherence but these larger algorithms require that coherence be maintained.
Thus the real challenge of porting a classical algorithm to a quantum computer is {\em removing all the implicit decoherence} present in the classical implementation, and in particular doing so {\em without incurring huge overheads}.

Karatsuba multiplication \cite{karatsuba1962multiplication}, the first multiplication algorithm with a sub-quadratic number of operations to be discovered, is an example of an algorithm where it is non-trivial to remove decoherence without sacrificing performance.
Karatsuba multiplication works by splitting its inputs $u, v$ into two halves $u=a+2^{h} b, v=x+2^{h} y$, recursively multiplying $ax$, $by$, and $(a+b)(x+y)$, then using those results to assemble the complete answer.
We demonstrate the basic idea in \fig{classical_code}, which has example python code for squaring a number using Karatsuba multiplication.
Note that this code has implicit decoherence.
In particular, it is allowing variables to go out of scope without being uncomputed.

\begin{figure}
\begin{python}
def classical_karatsuba_square(v: int) -> int:
    n = v.bit_length()
    if n <= 32:
        # Base case. Use schoolbook multiplication.
        return v**2

    pivot = n >> 1
    low = v & ~(-1 << pivot)
    high = v >> pivot

    low_sq = karatsuba_square(low)
    high_sq = karatsuba_square(high)
    sum_sq = karatsuba_square(low + high)

    return sum([
        low_sq,
        # 2ab = (a+b)**2 - a**2 - b**2
        (sum_sq - low_sq - high_sq) << pivot,
        high_sq << (pivot << 1),
    ])
\end{python}
\caption{
\label{fig:classical_code}
   Example python implementation of Karatsuba squaring.
   Works fine in a classical context, but contains implicit decoherence (e.g. the results of the recursive calls are discarded instead of uncomputed) that make it unsuitable in a quantum context.
}
\end{figure}

One way to remove the implicit decoherence in the code is to uncompute the recursive calls by running them in reverse.
This does remove the decoherence, but it increases the number of recursive calls from 3 to 6.
The recurrence relation that defines the cost of the code is $T(n) = O(n) + r \cdot T(n/2)$ where $T(1) = O(1)$ and $r$ is the number of recursive calls.
When $r>2$, the solution to this recurrence relation is $T(n) = O(n^{\lg r})$ where $\lg$ is the base-2 logarithm.
Increasing $r$ from 3 to 6 increases the total number of operations from sub-quadratic $O(n^{\lg 3})$ to cubic $O(n^3)$, which is clearly unacceptable.
Even schoolbook multiplication achieves $O(n^2)$ operations.

Another way to remove the decoherence is to store all intermediate values that would have been discarded during execution of the algorithm, and then uncompute those intermediate values in reverse order at the end of the algorithm \cite{bennett1973logical}.
This at most doubles the number of operations, but has the downside of increasing the space usage from $O(n)$ to the number of operations $O(n^{\lg 3})$.
It is possible to improve over this naive space bound by carefully analyzing the dependencies between values as a pebble game \cite{bennett1989pebble}.
Parent et al. do this in \cite{parent2017karatsuba}, and improve the space complexity to $O(n^{1.427})$ while preserving the time complexity.
It is unclear if it is possible to get all the way down to linear space cost using this approach.

In this paper, we present an alternative method for removing decoherence from Karatsuba multiplication.
Specifically, we orchestrate the execution of the algorithm such that intermediate values can simply be added directly into sections of the output register.
This avoids the need to store and uncompute the intermediate values, achieving $O(n)$ space usage while preserving the $O(n^{\lg 3})$ operation count, matching the asymptotic behavior of the classical implementation.

The paper is arranged as follows.
In \sec{introduction} we motivate the difficulty of the problem and discuss some background.
In \sec{methods} we describe how to rewrite Karatsuba multiplication into a series of recursive inline additions.
In \sec{results} we analyze the cost of our construction, reference our python and Q\# implementations, and present gate counts from Q\#'s tracing simulator.
Finally, \sec{conclusion} makes some closing observations.

\section{Construction}
\label{sec:methods}

\newcommand{\pluseq}{\mathrel{+}=}
\newcommand{\minuseq}{\mathrel{-}=}
\newcommand{\timeseq}{\mathrel{\ast}=}

Our method for transforming Karatsuba multiplication into a series of inline additions is a sequence of trivial rewrites.
In this section, we go over those rewrites one by one.

We start with a single instruction that performs the task we wish to achieve: offsetting some target $t$ by an amount equal to the product of two inputs $u \cdot v$.

$$t \pluseq u \cdot v$$

We assume that $u$ and $v$ are integers that have each been divided into $m$ words of size $w$, where $m$ is a power of 2.
If $m=1$ then we will perform a base case multiplication, e.g. schoolbook multiplication.
In the rest of this section we are interested in the case where $m>1$.

The next five rewrites we apply are determined by the fact that we wish to implement Karatsuba multiplication.
We split $u$ into two halves, $a$ and $b$, each with $h=m/2$ words.
We similarly split $v$ into two halves, $x$ and $y$.
We rewrite in terms of these new variables:

$$t \pluseq (a + b 2^{wh}) \cdot (x + y 2^{wh})$$

We then distribute, taking care to group values that will correspond to the recursive multiplications:

$$t \pluseq (ax) + (ay + bx) 2^{wh} + (by)4^{wh}$$

We rewrite the middle term to be in terms of $(a+b)(x + y)$, $ax$, and $by$:

$$t \pluseq (ax) + ((a+b)(x + y) - ax - by) 2^{wh} + (by)4^{wh}$$

We group terms corresponding to the same recursive call:

$$t \pluseq (ax) \cdot (1 - 2^{wh}) + (a+b)(x + y) \cdot 2^{wh} + (by) \cdot 2^{wh} \cdot (2^{wh}-1)$$

And we complete the transition to Karatsuba-style multiplication by putting each recursive half-sized multiplication on its own line:

$$\begin{aligned}
t &\pluseq (ax) \cdot (1 - 2^{wh})
\\
t &\minuseq (by) \cdot 2^{wh} \cdot (1-2^{wh})
\\
t &\pluseq (a+b)(x + y) \cdot 2^{wh}
\end{aligned}$$

Our goal now is to rewrite the recursive calls into a form where they are not scaled before being added into $t$.
We need to move the factors of $2^{wh}$ and $1-2^{wh}$ into separate instructions, away from the recursive terms $ax$, $by$, and $(a+b)(x+y)$.
This will allow us to perform an optimization analogous to tail-call optimization, where outputs go directly into the result instead of through a series of intermediates.

The multiplications by $2^{wh}$ are easily dealt with, because they correspond to word-aligned shifts.
We simply add into $t$ starting at word position $h$ instead of word position 0.
We indicate this by replacing $t$ with $t[h:]$ on the left hand side:

$$\begin{aligned}
t &\pluseq (ax) \cdot (1 - 2^{wh})
\\
t[h:] &\minuseq (by) \cdot (1 - 2^{wh})
\\
t[h:] &\pluseq (a+b)(x + y)
\end{aligned}$$

For the factors of $1-2^{wh}$, we use a slightly more complicated technique.
Instead of scaling the result being added into $t$, we temporarily inverse-scale $t$ while adding in the result.
We pre-multiply $t$ by the multiplicative inverse of $1 - 2^{wh}$, then perform the additions we want to be scaled, then post-multiply $t$ by $1 - 2^{wh}$.
This is equivalent to the original scaled addition:

$$\begin{aligned}
t &\timeseq  (1 - 2^{wh})^{-1}
\\
t &\pluseq ax
\\
t[h:] &\minuseq by
\\
t &\timeseq  1 - 2^{wh}
\\
t[h:] &\pluseq (a+b)(x + y)
\end{aligned}$$

Multiplying a number by $1-2^{wh}$ is equivalent to subtracting that number times $2^{wh}$ from itself.
This allows us to rewrite multiplications by this constant, and its multiplicative inverse, into a self-targeting addition.
This operation is well-defined despite the possible self-reference issues because, as we will discuss later, we pad the words so that additions into a word do not carry into the next word.
Because the input and output regions are aliased, the order in which we iterate over the word pairs being added is important:

$$\begin{aligned}
t[h:] &\pluseq t \text{  (using loop with increasing index)}
\\
t &\pluseq ax
\\
t[h:] &\minuseq by
\\
t[h:] &\minuseq t \text{  (using loop with decreasing index)}
\\
t[h:] &\pluseq (a+b)(x + y)
\end{aligned}$$

The last remaining intermediate values are the $a+b$ and $x+y$ expressions.
Technically it would be acceptable to temporarily allocate memory to hold these values during the recursion, but it is more space efficient to instead temporarily store them into $a$ and $x$ and so we do that:

$$\begin{aligned}
t[h:] &\pluseq t \text{  (using loop with increasing index)}
\\
t &\pluseq ax
\\
t[h:] &\minuseq by
\\
t[h:] &\minuseq t \text{  (using loop with decreasing index)}
\\
a &\pluseq b
\\
x &\pluseq y
\\
t[h:] &\pluseq ax
\\
a &\minuseq b
\\
x &\minuseq y
\end{aligned}$$

The final change we must make is to guarantee that, when we perform an operation like $t \pluseq ax$, it only affects $O(h)$ of the words in $t$.
In particular, it would be unacceptable for each of these additions to propagate a carry across the entire span of the register that $t$ is a subview of, because that would have a cost proportional to $O(n)$ instead of $O(wh)$.
We need a mechanism to terminate these carries.

The mechanism we use to terminate the carries is to pad the words in $t$ with enough bits to ensure they can directly store a product of words from $u$ and $v$.
Actually, because we add together words in $u$, $v$, and $t$, the padding needs to be sufficient to store a sum of products of sums of words from $u$ and $v$.
If we are dividing the input into $m$ words of size $w$, then each word in $t$ must have $2w + 3 \lg m$ bits of storage.
Similarly, because we are storing sums of words from $u$ and $v$ into the words of $u$ and $v$, the initial words in $u$ and $v$ must be padded up from $w$ bits to $w + \lg m$ bits.

Taking into account these considerations, we can scope the extent of every addition.
We indicate the scoping by using python slice notation.
The value $t[i:j]$ refers to the words in $t$ starting at index $i$ and continuing until just before index $j$:

\begin{equation}\label{eq:instructions}\begin{aligned}
t[h:4h] &\pluseq t[0:3h] \text{  (using loop with increasing index)}
\\
t[0:2h] &\pluseq ax
\\
t[h:3h] &\minuseq by
\\
t[h:4h] &\minuseq t[0:3h] \text{  (using loop with decreasing index)}
\\
a &\pluseq b
\\
x &\pluseq y
\\
t[h:3h] &\pluseq ax
\\
a &\minuseq b
\\
x &\minuseq y
\end{aligned}\end{equation}

This final series of instructions is the heart of our algorithm for Karatsuba multiplication without decoherence.
We iteratively decompose multiplications into this set of instructions until we hit the base case $m=1$, resulting in a long series of inline additions (and subtractions) and base case multiplications.
If the base case multiplication also decomposes into additions, then the whole algorithm is just a series of inline additions.

There is now only one remaining problem: once the multiplication has completed, we need to remove the padding bits.
This is challenging, because the result of the multiplication is spread in a disorganized fashion over the non-padding and padding bits of $t$.
We need to somehow merge the result onto just the non-padding bits, and uncompute the padding bits.

We fix this problem in a naive fashion.
During the initial preparation to perform a multiplication, we allocate a temporary register to play the role of $t$, instead of setting $t$ to be the actual target of the multiplication.
(To be clear, we do not do this during the recursive steps!
We only do it at the top-most level of the algorithm.)
We add the product $uv$ into the temporary register, with padding, using the recursive procedure we have been describing throughout the rest of this section.
We then iterate over groups of padding and non-padding bits in the temporary register, adding them into the true target at the appropriate offsets.
We then uncompute the temporary register.

Computing and uncomputing a temporary register doubles the cost of the multiplication, but it removes all remaining decoherence and completes our construction.

\section{Analysis}
\label{sec:results}

Given the word size $w$ and initial word count $m$, and recalling that we are operating on padded input words of size $w + \lg m$ and padded output words of size $2w + 3 \lg m$, we can determine the ultimate operation count of recursively rewriting multiplications into \eq{instructions}.
The resulting recurrence relation is

\begin{equation}
    \begin{aligned}
        T(1) &= O(B(w + \lg m))
        \\
        T(k) &= O(k \cdot (w + \lg m)) + 3 T(k/2)
    \end{aligned}
\end{equation}

Where $B(x)$ is the cost of a base-case multiplication of size $x$.
The solution to this recurrence relation is:

\begin{equation}
    \begin{aligned}
T(k) &= O(k (w + \lg m)) + 3 T(k/2)
\\   &= O\left(\left(\sum_{j=0}^{\lg k - 1} k (w + \lg m) \left(\frac{3}{2}\right)^j\right) + 3^{\lg k} B(w + \lg m)\right)
\\   &= O\left(k \cdot (w + \lg m) \left(\frac{3}{2}\right)^{\lg k} + 3^{\lg k} B(w + \lg m)\right)
    \end{aligned}
\end{equation}

If we set $w=\lg n$ and $m=n/\lg n$, where $n$ is the number of bits in each input number, then the total operation count is:

\begin{equation}
    \begin{aligned}
T(m) &= T\left(\frac{n}{\lg n}\right)
\\   &= O\left(\frac{n}{\lg n} \left(\lg n + \lg \frac{n}{\lg n}\right) \left(\frac{3}{2}\right)^{\lg \frac{n}{\lg n}} + 3^{\lg \frac{n}{\lg n}} B\left(\lg n + \lg \frac{n}{\lg n}\right)\right)
\\   &= O\left(n \left(\frac{3}{2}\right)^{\lg n} + 3^{\lg n - \lg \lg n} B(\lg n)\right)
\\   &= O\left(n^{\lg 3} + \frac{n^{\lg 3}}{\lg^{\lg 3} n} B(\lg n)\right)
\\   &= O\left(n^{\lg 3} \frac{B(\lg n)}{\lg^{\lg 3} n}\right)
    \end{aligned}
\end{equation}

From this analysis, we can see that it would not be asymptotically acceptable to use schoolbook multiplication in our base case because the resulting complexity would scale like $O(n^{\lg 3} \lg^{0.42} n)$ instead of $O(n^{\lg 3})$.
Fortunately, because our base case has size $w + \lg m = O(\lg n)$, we can simply use a space-inefficient form of Karatsuba multiplication, e.g. the construction from \cite{parent2017karatsuba}, as our base case.
The additive space cost we pay for this is polylogarithmic in $n$; asymptotically negligible.
This reduces the operation count to $O(n^{\lg 3})$ as desired.

We can also determine the space usage of the algorithm from our choices of $m$ and $w$.
The padded input has size $m \cdot (w + \lg m) = \frac{n}{\lg n} (\lg n + \lg \frac{n}{\lg n}) \leq 2 n$.
The padded output has size $2m \cdot (2w + 3 \lg m) = \frac{2n}{\lg n} (2 \lg n + 3 \lg \frac{n}{\lg n}) \leq 10 n$.
All other sources of memory usage, such as the base case multiplications, are negligible.
Thus the total space usage is $O(n)$ as desired.

Achieving a good asymptotic depth requires some tweaks to the construction.
The problem is that the recursive calls we are performing require exclusive access to overlapping ranges of the output register, which prevents the recursive cases from being run in parallel, which serializes the execution of the base cases.
We can work around this issue by making the base case larger.
If we set the word size $w^\prime$ to $n^{1 / \lg 3}$ instead of $\lg n$, then in the base case multiplication we can use the naive Karatsuba construction where all intermediate values are stored until the (base case) multiplication has completed (at which point the intermediate values are uncomputed in reverse order) \cite{bennett1973logical}.
We can afford to do this because the word size $w^\prime$ is small enough that the space used while executing a base case will be $O({w^\prime}^{\lg 3}) = O(n)$.
By using log-depth adders and parallel execution of smaller cases, the base case can be completed in $D(w^\prime) = O(\lg w^\prime) + D(w^\prime/2) = O(\lg^2 w^\prime) = O(\lg^2 n)$ depth while maintaining the desired $O({w^\prime}^{\lg 3}) = O(n)$ space usage (because the space usage cannot exceed the operation count).
There will be $3^{\lg n - \lg w^\prime} = O(n^{\lg 3 - 1})$ of these base cases, executed serially.
The depth cost of executing these base cases dominates the depth cost of the non-recursive additions used to prepare them (as long as those additions use log-depth adders).
Thus the total depth of this increased-word-size massively-parallel-base-case construction is $O(n^{\lg 3 - 1} \lg^2 n)$ and since the space usage is still linear the spacetime volume is $O(n^{\lg 3} \lg^2 n)$.
This is an improvement over the $O(n^{1 + \lg 3})$ volumes achieved in \cite{portugal2006karatsuba} and \cite{parent2017karatsuba}.

In order to double-check our algorithm and our analysis, we implemented our construction in both python and Q\#.
The python implementation is more performant, which allowed us to test larger cases.
The Q\# implementation allowed access to simulators that can compute explicit resource counts and verify we are not performing irreversible operations.
For simplicity, the code uses schoolbook multiplication as the base case multiplication (instead of a different Karatsuba construction).
This has no effect on the resulting data, because we do not simulate cases large enough for Karatsuba multiplication in the base case to outperform schoolbook multiplication.

\begin{figure}
    \centering
    \resizebox{\linewidth}{!}{
        \includegraphics{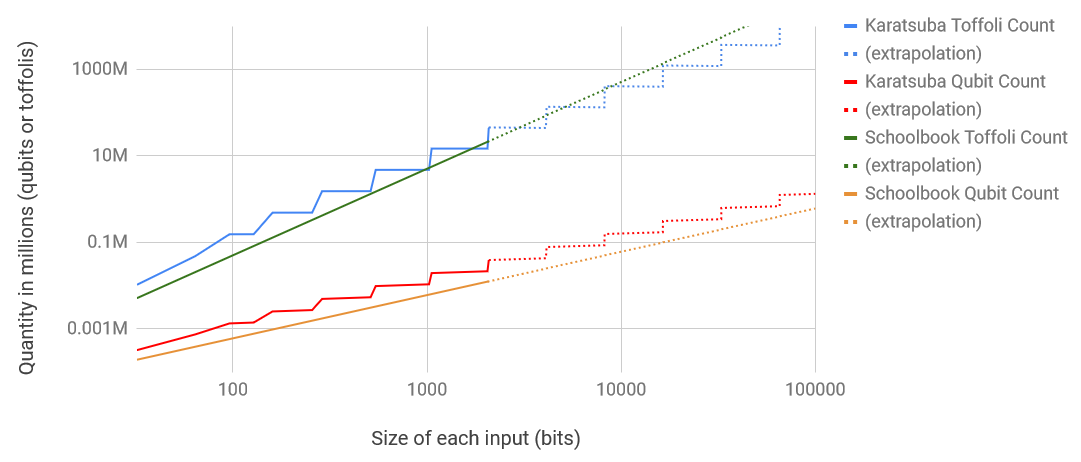}
    }
    \caption{
    \label{fig:data}
    Log-log plot of the number of Toffoli gates and qubits used by our Q\# implementation of Karatsuba multiplication and schoolbook multiplication for various input sizes.
    The staircase behavior in the Karatsuba curves are due to the implementation rounding to powers of 2.
    }
\end{figure}

Both code bases are included as ancillary files to this paper, and can be viewed online at \href{https://github.com/strilanc/quantum-karatsuba-2019}{github.com/strilanc/quantum-karatsuba-2019}.
Excerpts showing the recursive step, the most important part, are in the appendices.
\app{python} has the python excerpt, and \app{qsharp} has the equivalent Q\# excerpt.
The Toffoli count and qubit count, as determined by the Q\# trace simulator, is shown in \fig{data}.

\section{Conclusions}
\label{sec:conclusion}

In this paper, we removed decoherence from Karatsuba multiplication without increasing its space complexity or time complexity.
We did this by rewriting the recursive step of Karatsuba multiplication into a form where the recursive steps can directly add into subsections of the output.
This removed the need to store, and later uncompute, intermediate results, which was the main obstacle preventing the quantum implementation from matching the classical implementation in asymptotic cost.

We implemented and tested our algorithm, using Q\#'s trace simulator to get concrete resource counts.
Our Toffoli counts are broadly similar to the ones from \cite{parent2017karatsuba}.
It is notable that the crossover point where our implementation of Karatsuba multiplication becomes more efficient than our implementation of schoolbook multiplication (around 10000 bits) is larger than the size of modern RSA keys (2048 to 8192 bits), suggesting one would prefer to use schoolbook multiplication in Shor's algorithm in practice.
However, in this paper we focused on asymptotic arguments and did not attempt to optimize constant factors.
We also ignored important practical considerations, such as the cost of routing qubits towards each other in order for them to interact.
Also, the case we analyzed (multiplication of two quantum integers) is different from the case that occurs in Shor's algorithm (controlled modular multiplication of a quantum integer by a classical integer).
Therefore we do not draw any conclusions on the matter of whether Karatsuba multiplication would be useful in Shor's algorithm in practice.

We view the quantum technique of having recursive calls directly mutate sections of the output as being analogous to the classical technique of tail-call optimization.
In the same way that using tail calls is key to optimizing the space complexity of classical recursive algorithms, we believe that using inline-mutation calls is key to optimizing the space complexity of quantum recursive algorithms.
It is a basic tool, important to include in any quantum algorithm design toolbox.

\section{Acknowledgements}

We thank Hartmut Neven for creating an environment where this research was possible in the first place.
We thank Austin Fowler for reviewing a draft of this paper, and making suggestions that improved it.

\bibliographystyle{plainnat}
\bibliography{refs}

\appendix

\section{Python Code Excerpt}
\label{app:python}

\begin{python}
def _add_product_into_pieces(input_pieces1: List[IntBuf],
                             input_pieces2: List[IntBuf],
                             output_pieces: List[IntBuf],
                             sign: int):
    """Inline Karatsuba multiplication over the pieces.

    Note that the pieces must be large enough to hold
    intermediate results.
    """
    if not input_pieces1:
        return
    if len(input_pieces1) == 1:
        output_pieces[0] += (
            int(input_pieces1[0]) *
            int(input_pieces2[0]) *
            sign
        )
        return
    h = len(input_pieces1) >> 1

    # Input1 is logically split into two halves (a, b)
    #   such that a + 2**wh * b equals the input.
    # Input2 is logically split into two halves (x, y)
    #   such that x + 2**wh * y equals the input.

    # -----------------------------------
    # Perform
    #     out += a*x * (1-2**wh)
    #     out -= b*y * 2**wh * (1-2**wh)
    # -----------------------------------
    # Temporarily inverse-multiply the output by 1-2**wh,
    # so that the following two multiply-adds are scaled
    # by 1-2**wh.
    for i in range(h, len(output_pieces)):
        output_pieces[i] += output_pieces[i - h]
    # Recursive multiply-add for a*x.
    _add_product_into_pieces(
        input_pieces1=input_pieces1[:h],
        input_pieces2=input_pieces2[:h],
        output_pieces=output_pieces[:2*h],
        sign=sign)
    # Recursive multiply-subtract for b*y.
    _add_product_into_pieces(
        input_pieces1=input_pieces1[h:2*h],
        input_pieces2=input_pieces2[h:2*h],
        output_pieces=output_pieces[h:3*h],
        sign=-sign)
    # Multiply output by 1-2**wh, completing the scaling
    # of the previous two multiply-adds.
    for i in range(h, len(output_pieces))[::-1]:
        output_pieces[i] -= output_pieces[i - h]

    # -------------------------------
    # Perform
    #     out += (a+b)*(x+y) * 2**wh
    # -------------------------------
    # Temporarily store a+b over a and x+y over x.
    for i in range(h):
        input_pieces1[i] += input_pieces1[i + h]
        input_pieces2[i] += input_pieces2[i + h]
    # Recursive multiply-add for (a+b)*(x+y).
    _add_product_into_pieces(
        input_pieces1=input_pieces1[:h],
        input_pieces2=input_pieces2[:h],
        output_pieces=output_pieces[h:3*h],
        sign=sign)
    # Restore a and x.
    for i in range(h):
        input_pieces1[i] -= input_pieces1[i + h]
        input_pieces2[i] -= input_pieces2[i + h]
\end{python}

\section{Q\# Code Excerpt}

\label{app:qsharp}
\begin{qsharp}
operation _PlusEqualProductUsingKaratsubaOnPieces (
        output_pieces: LittleEndian[],
        input_pieces_1: LittleEndian[],
        input_pieces_2: LittleEndian[]) : Unit {
  body (...) {
    let n = Length(input_pieces_1);
    if (n <= 1) {
      if (n == 1) {
        PlusEqualProductUsingSchoolbook(
            output_pieces[0],
            input_pieces_1[0],
            input_pieces_2[0]);
      }
    } else {
      let h = n >>> 1;

      // Input 1 is logically split into two halves (a, b)
      //   such that a + 2**wh * b equals the input.
      // Input 2 is logically split into two halves (x, y)
      //   such that x + 2**wh * y equals the input.

      //-----------------------------------
      // Perform
      //     out += a*x * (1-2**wh)
      //     out -= b*y * 2**wh * (1-2**wh)
      //-----------------------------------
      // Temporarily inverse-multiply the output by 1-2**wh,
      // so that the following two multiplied additions are
      // scaled by 1-2**wh.
      for (i in h..Length(output_pieces) - 1) {
        PlusEqual(output_pieces[i], output_pieces[i - h]);
      }
      // Recursive multiply-add for a*x.
      _PlusEqualProductUsingKaratsubaOnPieces(
        output_pieces[0..2*h-1],
        input_pieces_1[0..h-1],
        input_pieces_2[0..h-1]);
      // Recursive multiply-subtract for b*y.
      Adjoint _PlusEqualProductUsingKaratsubaOnPieces(
        output_pieces[h..3*h-1],
        input_pieces_1[h..2*h-1],
        input_pieces_2[h..2*h-1]);
      // Multiply output by 1-2**wh, completing the scaling
      // of the previous two multiply-adds.
      for (i in Length(output_pieces) - 1..-1..h) {
        Adjoint PlusEqual(output_pieces[i],
                          output_pieces[i - h]);
      }

      //-------------------------------
      // Perform
      //     out += (a+b)*(x+y) * 2**wh
      //-------------------------------
      // Temporarily store a+b over a and x+y over x.
      for (i in 0..h-1) {
        PlusEqual(input_pieces_1[i],
                  input_pieces_1[i + h]);
        PlusEqual(input_pieces_2[i],
                  input_pieces_2[i + h]);
      }
      // Recursive multiply-add for (a+b)*(x+y).
      _PlusEqualProductUsingKaratsubaOnPieces(
        output_pieces[h..3*h-1],
        input_pieces_1[0..h-1],
        input_pieces_2[0..h-1]);
      // Restore a and x.
      for (i in 0..h-1) {
        Adjoint PlusEqual(input_pieces_1[i],
                          input_pieces_1[i + h]);
        Adjoint PlusEqual(input_pieces_2[i],
                          input_pieces_2[i + h]);
      }
    }
  }
  adjoint auto;
}
\end{qsharp}

\end{document}